# Similarity theory and calculation of turbulent fluxes at the surface for the stably stratified atmospheric boundary layers

by Sergej Zilitinkevich[1,2,3] and Igor Esau[2],

[1] Division of Atmospheric Sciences, University of Helsinki, Finland

[2] Nansen Environmental and Remote Sensing Centre / Bjerknes Centre for Climate Research, Bergen, Norway

[3] Finnish Meteorological Institute, Helsinki, Finalnd

# Similarity theory and calculation of turbulent fluxes at the surface for the stably stratified atmospheric boundary layers

by Sergej S. Zilitinkevich[1,2,3] and Igor N. Esau[2]

[1] Division of Atmospheric Sciences, University of Helsinki, Finland
[2] Nansen Environmental and Remote Sensing Centre / Bjerknes Centre for Climate Research, Bergen, Norway
[3] Finnish Meteorological Institute, Helsinki, Finland

**Abstract**

In this paper we revise the similarity theory for the stably stratified atmospheric boundary layer (ABL), formulate analytical approximations for the wind velocity and potential temperature profiles over the entire ABL, validate them against large-eddy simulation and observational data, and develop an improved surface flux calculation technique for use in operational models.

## 1. Introduction

Parameterisation of turbulence in atmospheric models comprise two basic problems:
- turbulence closure – to calculate vertical turbulent fluxes, first of all, the fluxes of momentum and potential temperature: $\vec{\tau}$ and $F_\theta$ through the mean gradients: $d\vec{U}/dz$ and $d\Theta/dz$ (where $z$ is the height, $\vec{U}$ and $\Theta$ are the mean wind speed and potential temperature);
- flux-profile relationships – to calculate the fluxes at the earth's surface: $u_*^2 = \tau_* = \tau|_{z=0}$ and $F_* = F_\theta|_{z=0}$ through the mean wind speed $U_1 = U|_{z=z_1}$ and potential temperature $\Theta_1 = \Theta|_{z=z_1}$ at a given level, $z_1$, above the surface.

This paper focuses on the flux-profile relationships for stable and neutral stratification. At first sight, it could be solved numerically using an adequate turbulence-closure model. However, this way is too computationally expensive: the mean gradients close to the surface are very sharp, which requires very high resolution, not to mention that the adequate closure for strongly stable stratification can hardly be considered as a fully understood, easy problem. Hence the practically sound problem is to analytically express the surface fluxes $\tau_*$ and $F_*$ through $U_1 = U|_{z=z_1}$ and $\Theta_1 = \Theta|_{z=z_1}$ available in numerical models (and similarly for the fluxes of humidity and other scalars). In numerical weather prediction (NWP) and climate models, the lower computational level is usually taken $z_1 \sim 30$ m (see Ayotte et al., 1996; Tjernstrom, 2004).

In neutral or near-neutral stratification the solution to the above problem is given by the logarithmic wall law:

$$\frac{dU}{dz} = \frac{\tau^{1/2}}{kz}, \quad \frac{d\Theta}{dz} = \frac{-F_\theta}{k_T \tau^{1/2} z}, \tag{1a}$$

$$U = \frac{\tau^{1/2}}{k} \ln \frac{z}{z_{0u}}, \quad \Theta - \Theta_s = \frac{-F_\theta}{k_T \tau^{1/2}} \ln \frac{z}{z_{0T}} = \Theta - \Theta_0 = \frac{-F_\theta}{k_T \tau^{1/2}} \ln \frac{z}{z_{0u}}, \tag{1b}$$

where $k$ and $k_T$ are the von Karman constants, $z_{0u}$ and $z_{0T}$ are the aerodynamic roughness lengths for momentum and heat, $\Theta_s$ is the potential temperature at the surface, and $\Theta_0$ is the aerodynamic surface potential temperature, that is the value of $\Theta(z)$ extrapolated logarithmically down to the level $z = z_{0u}$ [determination of the difference $\Theta_0 - \Theta_s = k_T^{-1}(-F_\theta \tau^{-1/2}) \ln(z_{0u}/z_{0T})$ comprises an independent problem; see, e.g., Zilitinkevich et al. (2001, 2002)]. As follows from Eq. (1), $\tau_1^{1/2} = kU_1 (\ln z/z_{0u})^{-1}$ and $F_{\theta 1} = -kk_T U_1 (\Theta_1 - \Theta_0)(\ln z/z_{0u})^{-2}$. The turbulent fluxes $\tau_1$ and $F_{\theta 1}$ at the level $z = z_1$ can be identified with the surface fluxes: $\tau_1 = \tau_*$ and $F_{\theta 1} = F_*$, provided that $z_1$ is much less then the height, $h$, of the atmospheric boundary layer (ABL). In neutral stratification, typical value of $h$ is a few hundred metres, so that the requirement $z_1 \approx 30$ m $<< h$ is satisfied.

In stable stratification, the problem becomes more complicated. Its commonly accepted solution is based, firstly, on the assumption that the level $z_1$ belongs to the surface layer [that is the lower one tenth of the ABL, where the turbulent fluxes do not diverge considerably from their surface values $\tau \approx \tau_*$ and $F_\theta \approx F_*$] and, secondly, on the Obukhov (MO) similarity theory for the surface-layer turbulence (Monin and Obukhov (1954).

The MO theory states that the turbulent regime in the stratified surface layer is fully characterized by the turbulent fluxes, $\tau \approx \tau_* = u_*^2$ and $F_\theta \approx F_*$, and the buoyancy parameter, $\beta = g/T_0$ (where $g$ is the acceleration of gravity, and $T_0$ is a reference value of absolute temperature), which determine the familiar MO length scale

$$L = \frac{\tau^{3/2}}{-\beta F_\theta}; \tag{2}$$

whereas the velocity and potential temperature gradients are expressed through universal functions, $\Phi_M$ and $\Phi_H$, of the dimensionless height $\xi = z/L$:

$$\frac{kz}{\tau^{1/2}} \frac{dU}{dz} = \Phi_M(\xi), \tag{3a}$$

$$\frac{k_T \tau^{1/2} z}{F_\theta} \frac{d\Theta}{dz} = \Phi_H(\xi). \tag{3b}$$

From the requirement of consistency with the wall law for the neutral stratification, Eq. (1), it follows that $\Phi_M = \Phi_H = 1$ at $\xi << 1$. The asymptotic behaviour of $\Phi_M$ and $\Phi_H$ in strongly stable stratification (at $\xi >> 1$) is traditionally derived from the concept of z-less stratification, which states that at $z >> L$ the distance over the surface, $z$, no longer affects turbulence. If so, $z$ should drop out from the velocity- and temperature-gradient formulations, which immediately suggests the linear asymptotes: $\Phi_M \sim \Phi_H \sim \xi$. The linear interpolation between the neutral and the strong stability limits gives

$$\Phi_M = 1 + C_{U1}\xi, \tag{4a}$$

$$\Phi_H = 1 + C_{\Theta 1}\xi, \tag{4b}$$

where $C_{U1}$ and $C_{\Theta 1}$ are empirical dimensionless constants.

The above analysis is usually considered as relevant only to the surface layer. However, the basic statement of the MO similarity theory, namely, that the surface-layer turbulence is fully characterised by $\tau$, $F_\theta$ and $\beta$, is applicable to locally generated turbulence in more general context. Nieuwstadt (1984) was probably the first who extended the MO theory by substituting the height-dependent $\tau$ and $F_\theta$ for the height-constant $\tau_*$ and $F_*$, and demonstrated its successful application to the entire nocturnal stable ABL. In the present paper we employ this extended version of the MO theory.

In the surface layer, substituting Eq. (4) for $\Phi_M$ and $\Phi_H$ into Eq. (3) and integrating over $z$ yields the log-linear approximation:

$$U = \frac{u_*}{k}\left(\ln\frac{z}{z_{u0}} + C_{U1}\frac{z}{L_s}\right), \tag{5a}$$

$$\Theta - \Theta_0 = \frac{-F_*}{k_T u_*}\left(\ln\frac{z}{z_{u0}} + C_{\Theta 1}\frac{z}{L_s}\right), \tag{5b}$$

where $L_s = u_*^3(-\beta F_*)^{-1}$.

Since the late fifties, Eqs. (3)-(5) have been compared with experimental data in numerous papers that basically gave estimates of $C_{U1}$ more or less close to 2 and $C_{\Theta 1}$ also close to 2 but with a wider spread (see overview by Högström, 1996). Experimentalists often admitted that for $\Theta$ the log-linear formulation works worse then for $U$, but no commonly accepted alternative formulations were derived from physical grounds. Esau and Byrkjedal (2006) analysed data from large-eddy simulations (LES) and disclosed that the coefficient $C_{\Theta 1}$ in Eq. (4.b) is not a constant but increases with increasing $z/L$.

According to Eqs. (3)-(4) the Richardson number, $\text{Ri} \equiv \beta(d\Theta/dz)(dU/dz)^{-2}$, monotonically increases with increasing $z/L$, and at $z/L \to \infty$ achieves its maximal value: $\text{Ri}_c = k^2 C_{\Theta 1} k_T^{-1} C_{U1}^{-2}$. In other words, Eq. (4) are not applicable to $\text{Ri} > \text{Ri}_c$. This conclusion is consistent with the critical Richardson number concept, universally accepted at the time when the MO theory and Eqs. (3)-(5) were formulated.

However, as recognised recently, the concept of the critical Ri contradicts to both experimental evidence and analysis of the turbulent kinetic and potential energy budgets. This conclusion is by no means new. Long ago it has been understood that turbulent closures or surface flux schemes implying the critical Ri lead to erroneous conclusions, such as unrealistic decoupling of air flows from underlying surface in all cases when Ri > $Ri_c$. It is not surprising that modellers do not use Eq. (4) as well as other formulations of similar type, even though they are supported by experimental data. Instead, operational modellers develop their own flux-profile relationships, free of Ri critical, and evaluate them indirectly – fitting the model results to the available observational data. Different points of view of experimentalists and operational modellers on the flux-profile relationships have factually caused two nearly independent lines of inquiry in this field (see discussion in Zilitinkevich et al., 2002).

One more point deserves emphasising. Currently used flux-calculation schemes identify the turbulent fluxes calculated at the level $z_1$ with the surface fluxes. However, in strongly stable stratification, especially in long-lived stable ABLs, the ABL height, $h$, quite often reduces to a few dozen metres[1] (see Zilitinkevich and Esau, 2002, 2003; Zilitinkevich et al., 2006a) and becomes comparable with $z_1$ adopted in operational models. In such cases $\tau_1$ and $F_{\theta 1}$ have nothing in common with $\tau_*$ and $F_*$.

Furthermore, the MO theory, considered over half a century as an ultimate background for analysing the surface layer turbulence, is now revised. Zilitinkevich and Esau (2005) have found that, besides $L$, Eq. (2), which characterise the stabilising effect of local buoyancy forces on turbulence, there are two additional length scales: $L_f$ characterising the effect of the Earth's rotation and $L_N$ characterizing the non-local effect of the static stability in the free atmosphere above the ABL:

$$L_N = \frac{\tau^{1/2}}{N}, \tag{6a}$$

$$L_f = \frac{\tau^{1/2}}{|f|}, \tag{6b}$$

where $N$ is the Brunt-Väisälä frequency at $z > h$ (typically $N \sim 10^{-2}$ s$^{-1}$), and $f$ is the Coriolis parameter. Interpolating between the squared reciprocals of the three scales (to give priority to stronger mechanisms that is to smaller scales) a composite turbulent length scale becomes:

$$\frac{1}{L_*} = \left[\left(\frac{1}{L}\right)^2 + \left(\frac{C_N}{L_N}\right)^2 + \left(\frac{C_f}{L_f}\right)^2\right]^{1/2}, \tag{7}$$

where $C_N = 0.1$ and $C_f = 1$ are empirical dimensionless coefficients[2]. Advantages of this scaling have been demonstrated in the plots of $\Phi_M$ and $\Phi_H$ versus $z/L_*$ (Figures 2 and 5 in *op. cit.*) showing essential collapse of data points compared to the traditional plots of $\Phi_M$ and $\Phi_H$ versus $z/L$.

---

[1] The ABL height is defined as the level at which the turbulent fluxes become an order of magnitude smaller than close to the surface.

[2] In *op. cit.* the coefficient $C_N$ was taken $0.1$ for $\Phi_M$ and $0.15$ for $\Phi_H$. Further analysis has shown that the difference is insignificant, which allows employing one composite length scale given by Eq. (7).

Practical application of this scaling requires information about vertical profiles of turbulent fluxes across the ABL. As demonstrated by Lenshow et al. (1988), Sorbjan (1988), Wittich (1991), Zilitinkevich and Esau (2005) and Esau and Byrkjedal (2006), the ratios $\tau/\tau_*$ and $F_\theta/F_*$ are reasonable accurately approximated by universal functions of $z/h$, where $h$ is the ABL height [see Eq. (15) below].

As follows from the above discussion, currently used surface flux calculation schemes need to be improved accounting for
- modern experimental evidence and theoretical developments arguing against the critical Ri concept,
- additional mechanisms and scales, first of all $L_N$, disregarded in the classical similarity theory for stable ABLs,
- essential difference between the surface fluxes and the fluxes at $z = z_1$.

In the present paper we attempt to develop a new scheme applicable to as wide as possible range of stable and neutral ABL regimes using recent theoretical developments and new, high quality data from observations and LES.

## 2. Mean gradients and Richardson numbers

Until recently the ABLs were distinguished accounting for only one factor, the potential temperature flux at the surface, $F_*$: neutral ABLs at $F_* = 0$, and stable ABLs at $F_* < 0$. Accounting for the recently disclosed role of the static stability above the ABL, we now apply more detailed classification:
- truly neutral (TN) ABL: $F_* = 0$, $N = 0$,
- conventionally neutral (CN) ABL: $F_* = 0$, $N > 0$,
- nocturnal stable (NS) ABL: $F_* < 0$, $N = 0$,
- long-lived stable (LS) ABL: $F_* < 0$, $N > 0$.

Realistic surface flux calculation scheme should be based on a model applicable to all these types of the ABL.

As mentioned in Section 1, Eq. (4b) gives erroneous asymptotic behaviour at large $\xi = z/L$ and leads to appearance of the critical Ri. This conclusion is sometimes treaded as a failure of the MO theory, but this is not the case. The MO theory states only that $\Phi_M$ and $\Phi_H$ are universal functions of $\xi$, whereas the linear forms of the $\Phi$-functions, Eq. (4), are derived form the heuristic concept of $z$-less stratification, which is neither proved theoretically nor confirmed by experimental data.

In fact, this concept is not needed to derive the linear asymptotic formula for the velocity gradient in the stationary, homogeneous, sheared flows in very strong static stability. Recall that the flux Richardson number is defined as the ratio of the consumption of turbulent kinetic energy (TKE) caused by the negative buoyancy forces, $-\beta F_\theta$, to the shear generation of the TKE, $\tau\, dU/dz$:

$$\mathrm{Ri}_f = \frac{-\beta F_\theta}{\tau dU/dz}. \tag{8}$$

$\mathrm{Ri}_f$ (in contrast to the gradient Richardson number, Ri) cannot grow infinitely: otherwise the TKE consumption would exceed its production. Hence $\mathrm{Ri}_f$ at very large $\xi$ should tend to a limit, $\mathrm{Ri}_f^\infty$ ($=0.2$ according to currently available experimental data). Then resolving Eq. (8) for $dU/dz$ and substituting $\mathrm{Ri}_f^\infty$ for $\mathrm{Ri}_f$ gives the asymptote

$$\frac{dU}{dz} \to \frac{\tau^{1/2}}{\mathrm{Ri}_f^\infty L}, \tag{9}$$

which in turn gives $\Phi_M \to k(\mathrm{Ri}_f^\infty)^{-1}\xi$, and thus rehabilitates Eq. (4) for $\Phi_M$. The gradient Richardson number becomes

$$\mathrm{Ri} \equiv \frac{\beta d\Theta/dz}{(dU/dz)^2} = \frac{k^2}{k_T}\frac{\xi \Phi_H(\xi)}{(1+C_{U1}\xi)^2}. \tag{10}$$

Therefore to assure unlimited growth of Ri with increasing $\xi$ (in other words, to guarantee "no Ri critical"), the asymptotic $\xi$-dependence of $\Phi_H$ should be stronger then linear. Recalling that the function $\Phi_H$ at small $\xi$ is known to be close to linear, a reasonable compromise could be quadratic polynomial [recall the above quoted conclusion of Esau and Byrkjedal (2006) that $C_{\Theta 1}$ in Eq. (4b) increases with increasing $z/L$).

On these grounds we adopt the approximations $\Phi_M = 1+C_{U1}\xi$ and $\Phi_H = 1+C_{\Theta 1}\xi + C_{\Theta 2}\xi^2$ covering the TN and NS ABLs. To extend them to the CN and LS ABLs, we employ the generalised scaling, Eqs. (6)-(7):

$$\Phi_M = 1+C_{U1}\frac{z}{L_*}, \tag{11a}$$

$$\Phi_H = 1+C_{\Theta 1}\frac{z}{L_*}\xi + C_{\Theta 2}\left(\frac{z}{L_*}\right)^2. \tag{11b}$$

Comparing Eqs. (9) and (11a) gives $\mathrm{Ri}_f^\infty = kC_{U1}^{-1}$. Then taking conventional values of $\mathrm{Ri}_f^\infty = 0.2$ and $k=0.4$ gives an "a prior" estimate of $C_{U1} = 2$.

Figures 1 and 2 show $\Phi_M$ and $\Phi_H$ versus $\xi = z/L_*$ after LES DATABASE64 (Beare et al., 2006; Esau and Zilitinkevich, 2006), which includes the TN, CN, NS, and LS ABLs. Figure 2 confirms that the $\xi$-dependence of $\Phi_H$ is indeed essentially stronger then linear: With increasing $\xi$, the best-fit linear dependence $\Phi_H = 1+2\xi$ shown by thin line diverge from data more and more, and at $\xi >> 1$ becomes unacceptable. The steeper thin line shows the quadratic asymptote $\Phi_H = 0.2\xi^2$ relevant only to very large $\xi$. Figure 1 confirms the expected linear dependence. Both figures demonstrate reasonably good performance of Eq.

(11) over the entire ABL depth (data for $z < h$ are indicated by dark grey points) and allow determining the constants $C_{U1} = 2$ (coinciding with the above "*a priori*" estimate), $C_{\Theta1} = 1.6$ and $C_{\Theta2} = 0.2$, with the traditional values of the von Karman constants: $k = 0.4$ and $k_T = 0.47$. For comparison, data for $z > h$ (indicated by light grey points) quite expectedly exhibit wide spread. The composite scale $L_*$ is calculated after Eqs. (6)-(7) with $C_N = 0.1$ and $C_f = 1$.

Figure 3 shows the gradient Richardson number, Eq. (10), versus $\xi = z / L$ after the LES data for TN and NS ABLs (indicated by dark and light grey points, as in Figures 1 and 2) and data from meteorological mast measurements at about 5, 9 and 13 m above the snow surface in the field campaign SHEBA (Uttal et al., 2002) indicated by green points. The bold curve shows our approximation of $\mathrm{Ri} = k^2 k_T^{-1} \xi \Phi_H \Phi_M^{-2}$ – taking $\Phi_M$ and $\Phi_H$ after Eq. (11) with $C_{U1} = 2$, $C_{\Theta1} = 1.6$ and $C_{\Theta2} = 0.2$; the thin curve shows the traditional approximation of Ri – taking $\Phi_M$ and $\Phi_H$ after Eq. (4) with $C_{U1} = 2$ and $C_{\Theta1} = 2$ (it affords critical value of $\mathrm{Ri} \approx 0.17$); and the steep thin line shows the asymptotic behaviour of our approximation, $\mathrm{Ri} \sim \xi$, at large $\xi$. Heavy points with error bars are the bin averaged values after LES DATABASE64.

This figure demonstrates consistency between the LES and the field data for so sensitive parameter as Ri (the ratio of gradients – inevitably determined with pronounced errors). For our analysis this result is critically important. It allows using the LES DATABASE64 on equal grounds with experimental data. Recall that using LES we have the advantage of fully controlled conditions, which is practically unachievable in field experiments.

We give here one example: Dealing with LES data we receive a possibility to distinguish between data for the ABL interior, $z < h$ (indicated in our figures by dark grey points) and data for $z > h$ (indicated by light grey points). As seen in Figure 3, the gradient Richardson number within the ABLs practically never exceeds 0.25 – 0.3, although turbulence is observed at much larger Ri. This observation perfectly correlates with recent theoretical conclusion that $\mathrm{Ri} \sim 0.25$ is not the critical Ri in old sense (the border between turbulent and laminar regimes) but a threshold separating the two turbulent regimes of essentially different nature: strong, chaotic turbulence at $\mathrm{Ri} \ll 0.25$; and weak, intermittent turbulence at $\mathrm{Ri} \gg 0.25$. These two are just the regimes typical of the ABLs and the fee atmosphere, respectively.

## 3. Surface fluxes

The above analysis clarifies our understanding of the physical nature of stable ABLs but does not immediately give flux-profile relationships suitable for practical applications. To receive analytical approximations of the mean wind and temperature profiles, $U(z)$ and $\Theta(z)$, across the ABL, we apply the generalised similarity theory presented in Section 2 to "characteristic functions":

$$\Psi_U = \frac{kU(z)}{\tau^{1/2}} - \ln\frac{z}{z_{0u}}, \tag{12a}$$

$$\Psi_\Theta = \frac{k_T \tau^{1/2}[\Theta(z) - \Theta_0]}{-F_\theta} - \ln\frac{z}{z_{0u}}, \tag{12b}$$

and employ LES DATABASE64 to determine their dependences on $\xi = z/L_*$.

Results from this analysis presented in Figures 4 and 5 are quite constructive. Over the entire ABL depth, $\Psi_U$ and $\Psi_\Theta$ show practically universal dependences on $\xi$ that can be reasonably accurately approximated by the power laws:

$$\Psi_U = C_U \xi^{5/6}, \tag{13a}$$

$$\Psi_\Theta = C_\Theta \xi^{4/5}, \tag{13b}$$

with $C_U = 3.0$ and $C_\Theta = 2.5$.

The wind and temperature profiles becomes

$$\frac{kU}{\tau^{1/2}} = \ln\frac{z}{z_{0u}} + C_U \left(\frac{z}{L}\right)^{5/6} \left[1 + \frac{(C_N N)^2 + (C_f f)^2}{\tau} L^2\right]^{5/12}, \tag{14a}$$

$$\frac{k_T \tau^{1/2}(\Theta - \Theta_0)}{-F_\theta} = \ln\frac{z}{z_{0u}} + C_\Theta \left(\frac{z}{L}\right)^{4/5} \left[1 + \frac{(C_N N)^2 + (C_f f)^2}{\tau} L^2\right]^{2/5}, \tag{14b}$$

where $C_N = 0.1$ and $C_f = 1$ [see discussion of Eq. (7)]. Given $U(z)$, $\Theta(z)$ and $N$, Eqs. (14a,b) allow determining the turbulent fluxes, $\tau$ and $F_\theta$, and the MO length, $L = \tau^{3/2}(-\beta F_\theta)^{-1}$, at the computational level $z$. Numerical solution to this system is simplified by the fact that the major terms on the right hand sides are the logarithmic ones, and moreover, the second terms in square brackets are usually small compared to unity. Hence iteration methods should work efficiently. As a first approximation $N$, unknown until we determine the ABL height, is taken $N = 0$. In the next iterations, it is calculated after Eq. (18).

Given $\tau$ and $F_\theta$, the surface fluxes are calculated using quasi-universal dependencies:

$$\frac{\tau}{\tau_*} = \exp\left[-\frac{8}{3}\left(\frac{z}{h}\right)^2\right], \tag{15a}$$

$$\frac{F_\theta}{F_*} = \exp\left[-2\left(\frac{z}{h}\right)^2\right]; \tag{15b}$$

for details see Zilitinkevich and Esau (2005) and Esau and Byrkjedal (2006).

The ABL height, $h$, required in Eq. (15) is calculated using the multi-limit $h$-model (Zilitinkevich et al., 2006a, and references therein) consistent with the present analyses. The diagnostic version of this model determines the equilibrium ABL height, $h_E$:

$$\frac{1}{h_E^2} = \frac{f^2}{C_R^2 \tau_*} + \frac{N|f|}{C_{CN}^2 \tau_*} + \frac{|f\beta F_*|}{C_{NS}^2 \tau_*^2}, \tag{16}$$

where $C_R = 0.6$, $C_{CN} = 1.36$ and $C_{NS} = 0.51$ are empirical dimensionless constants.

More accurately $h$ can be calculated using prognostic, relaxation equation (Zilitinkevich and Baklanov, 2002):

$$\frac{\partial h}{\partial t} + \vec{U} \cdot \nabla h - w_h = K_h \nabla^2 h - C_t \frac{u_*}{h_E}(h - h_E),  \qquad (17)$$

which therefore should be incorporated in a numerical model employing our scheme. In Eq. (17), $h_E$ is taken after Eq. (16), $w_h$ is the mean vertical velocity at the height $z = h$ (available in numerical models), the combination $C_t u_* h_E^{-1}$ expresses the relaxation time scale, $C_t \approx 1$ is an empirical dimensionless constant, and $K_h$ is the horizontal turbulent diffusivity (same as in other prognostic equations of the model under consideration).

Finally, given $h$, the free-flow Brunt-Väisälä frequency, $N$, is determined through the root mean square value of the potential temperature gradient over the layer $h < z < 2h$:

$$N^4 = \frac{1}{h} \int_h^{2h} \left( \beta \frac{\partial \Theta}{\partial z} \right)^2 dz \qquad (18)$$

and substituted into Eq. (14) for the next iteration.

Some problems (first of all, air-sea interaction) require not only the absolute value of the surface momentum flux, $\vec{\tau}_*$, but also its direction. Recalling that our method allows determining the ABL height, $h$, and therefore the wind vector at this height, $\vec{U}_h$, the problem reduces to the determination of the angle, $\alpha_*$ between $\vec{U}_h$ and $\vec{\tau}_*$. For this purpose we employ the cross-isobaric angle formulation:

$$\sin \alpha_* = \frac{-fh}{kU_h} \left[ -2 + 10 \frac{(-\beta F_* h)^2}{\tau_*^3} + 0.225 \frac{(Nh)^2}{\tau_*} + 10 \frac{(fh)^2}{\tau_*} \right], \qquad (19)$$

based on the same similarity theory as the preset paper [see Eqs. (7b), (41b), (43) and Figure 4 in Zilitinkevich and Esau (2005)].

Following the above procedure, Eqs. (14)-(18) allow calculating the following parametres:
- turbulent fluxes $\tau$ (z) and $F_\theta$ (z) at any computational level $z$ within the ABL,
- surface fluxes, $\vec{\tau}_*$ and $F_*$,
- ABL height, $h$, [either diagnostically after Eq. (16) or more accurately, accounting for it evolution after Eqs. (16)-(17)].

Empirical constants that appear in the above formulations are given in Table 1.

The proposed method can be applied, in particular, to shallow ABLs, when the lowest computational level is close to $h$, and standard approach completely fails. But it has advantages also in usual situations when the ABL (the height interval $0 < z < h$) contains several computational levels. In such cases, it provides several independent estimates of $h$, $u_*^2$ and $F_*$, and by this means makes available a kind of data assimilation, namely, more reliable determination of $h$, $u_*^2$ and $F_*$ through averaging over all estimates.

## 4. Concluding remarks

In this paper we employ a generalised similarity theory for the stably stratified sheared flows accounting for non-local features of atmospheric stable ABLs, follow modern views on the turbulent energy transformations rejecting the critical Richardson number concept, and use recent, high quality experimental and LES data to develop analytical formulations for the wind velocity and potential temperature profiles across the entire ABL.

Results from our analysis are validated using LES data from DATABASE64 covering the four types of ABLs: truly neutral, conventionally neutral, nocturnal stable and long-lived stable. These LES are in turn validated through (shown to be consistent with) observational data from the field campaign SHEBA.

Employing generalised format for the dimensionless velocity and potential temperature gradients, $\Phi_M$ and $\Phi_H$, Eq. (3), based on the composite turbulent length scale $L_*$, Eq. (7), and $z$-dependent turbulent velocity and temperature scales, $\tau^{1/2}$ and $F_\theta \tau^{-1/2}$, we demonstrate that $\Phi_M$ and $\Phi_H$ are to a reasonable accuracy approximated by universal functions of $z/L_*$ ($\Phi_M$ linear, $\Phi_H$ stronger then linear) across the entire ABL.

Using the quadratic polynomial approximation for $\Phi_H$, we demonstrate that our formulation leads to the unlimitedly increasing $z/L$-dependence of the gradient Richardson number, Ri, consistent with both LES and field data and arguing against the critical Ri concept.

We employ the above generalised format to the deviations, $\Psi_U$ and $\Psi_\Theta$, Eq. (12), of the dimensionless mean wind and potential temperature profiles from their logarithmic parts [ ~ $\ln(z/z_{0u})$ ] to obtain power-law approximations: $\Psi_U \sim (z/L_*)^{5/6}$ and $\Psi_\Theta \sim (z/L_*)^{4/5}$ that perform quite well across the entire ABL.

On this basis, employing also our prior ABL height model and resistance laws, we propose a new method for calculating the turbulent fluxes at the surface in numerical models.

## Acknowledgements

This work has been supported by the EU Marie Curie Chair Project MEXC-CT-2003-509742, ARO Project W911NF-05-1-0055, EU Project FUMAPEX EVK4-2001-00281, Norwegian Project MACESIZ 155945/700, joint Norway-USA Project ROLARC 151456/720, and NORDPLUS Neighbour 2006-2007 Project 177039/V11.

# References

Ayotte, K. W., Sullivan, P. P., Andren, A., Doney, S. C., Holtslag, A. A. M., Large, W. G., McWilliams, J. C., Moeng, C.-H., Otte, M., Tribbia, J. J., and Wyngaard, J., 1996: An evaluation of neutral and convective planetary boundary-layer parameterizations relative to large eddy simulations, *Boundary-Layer Meteorol.*, **79**, 131-175.

Beare, R. J., MacVean, M. K., Holtslag, A. A. M., Cuxart, J., Esau, I., Golaz, J. C., Jimenez, M. A., Khairoudinov, M., Kosovic, B., Lewellen, D., Lund, T. S., Lundquist, J. K., McCabe, A., Moene, A. F., Noh, Y., Raasch, S., and Sullivan, P., 2006: An intercomparison of large eddy simulations of the stable boundary layer, *Boundary Layer Meteorol.* **118**, 247 – 272.

Esau, I., and Byrkjedal, Ø, 2006: Application of large eddy simulation database to optimization of first order closures for neutral and stably stratified boundary layers, arXiv preprint available on http://arxiv.org/abs/physics/0612206.

Esau, I. N., and Zilitinkevich, S. S., 2006: Universal dependences between turbulent and mean flow parametres in stably and neutrally stratified planetary boundary layers. *Nonlin. Processes Geophys.*, **13**, 135–144.

Högström, U., 1996: Review of some basic characteristics of the atmospheric surface layer, *Bound-Layer Meteorol.*, **78**, 215–246.

Lenschow, D. H., Li, X. S., Zhu, C. J., and Stankov, B. B., 1988: The stably stratified boundary layer over the Great Plains. Part 1: Mean and turbulence structure. *Boundary-layer Meteorol.*, **42**, 95-121.

Monin, A. S., and Obukhov, A. M., 1954: Main characteristics of the turbulent mixing in the atmospheric surface layer, *Trudy Geophys. Inst. AN. SSSR,* 24(151), 153-187.

Nieuwstadt, F. T. M., 1984: The turbulent structure of the stable, nocturnal boundary layer, *J. Atmos. Sci.*, **41**, 2202-2216.

Sorbjan, Z., 1988: Structure of the stably-stratified boundary layer during the SESAME-1979 experiment. *Boundary-Layer Meteorol.*, **44**, 255-266.

Tjernstrom, M., Zagar, M., Svensson, G., Cassano, J. J., Pfeifer, S., Rinke, A., Wyser, A., Dethloff, K., Jones, C., Semmler, T., and M. Shaw, 2004: Modelling the arctic boundary layer: an evaluation of six ARCMIP regional-scale models using data from the SHEBA project. *Boundary-Layer Meteorol.*, **117**, 337–381.

Uttal, T., and 26 co-authors, 2002: Surface Heat Budget of the Arctic Ocean. *Bull. Amer. Meteorol. Soc.* **83**, 255 – 275.

Wittich, K. P., 1991: The nocturnal boundary layer over Northern Germany: an observational study. *Boundary-Layer Meteorol.*, **55**, 47-66.

Yague, C., Viana, S., Maqueda G., and Redondo, J. M., 2006: Influence of stability on the flux-profile relationships for wind speed, phi-m, and temperature, phi-h, for the stable atmospheric boundary layer, *Nonlin. Processes Geophys.*, **13**, 185–203

Zilitinkevich, S. S., and Baklanov, A., 2002: Calculation of the height of stable boundary layers in practical applications. *Boundary-Layer Meteorol.* **105**, 389-409.

Zilitinkevich S. S., and Esau, I. N., 2002: On integral measures of the neutral, barotropic planetary boundary layers. *Boundary-Layer Meteorol.* **104**, 371-379.

Zilitinkevich S. S. and Esau I. N., 2003: The effect of baroclinicity on the depth of neutral and stable planetary boundary layers. *Quart, J. Roy. Meteorol. Soc*. **129**, 3339-3356.

Zilitinkevich, S. S. and I. Esau, 2005: Resistance and heat transfer laws for stable and neutral planetary boundary layers: old theory, advanced and ee-evaluated, *Quart. J. Roy. Meteorol. Soc.*, **131**, 1863-1892.

Zilitinkevich, S. S., Grachev, A. A., and Fairall, C. W., 2001: Scaling reasoning and field data on the sea-surface roughness lengths for scalars. *J. Atmos. Sci.*, **58**, 320-325.

Zilitinkevich, S. S., Perov, V. L., King, J. C., 2002: near-surface turbulent fluxes in stable stratification: calculation techniques for use in general circulation models. *Qurt. J. Roy. Meteorol. Soc.*, **128**, 1571-1587.

Zilitinkevich, S. S., Esau, I., and Baklanov, A., 2006a: Further comments on the equilibrium height of neutral and stable planetary boundary layers. *Qurt. J. Roy. Meteorol. Soc.*, In press.

**Table 1**

| Constant | in Equation | Comments |
|---|---|---|
| $k = 0.4$, $k_T = 0.47$ | (1), (3), etc | traditional values |
| $C_N = 0.1$, $C_f = 1$ | (7) | after Zilitinkevich and Esau (2005), slightly corrected |
| $C_{U1} = 2.0$, $C_{\Theta 1} = 1.6$, $C_{\Theta 2} = 0.2$ | (11a,b) | after present paper; $C_{U1} = 2.0$ and $C_{\Theta 1} = 1.6$ correspond to the coefficients $\beta_1 = C_{U1}/k = 5.0$ and $\beta_2 = C_{\Theta 1}/k = 4.0$ in the log-linear laws formulated for $L = u_*^3(-k\beta F_*)^{-1}$ |
| $C_U = 3.0$, $C_\Theta = 2.5$ | (13), (14) | after present paper |
| $C_R = 0.6$, $C_{CN} = 1.36$, $C_{NS} = 0.51$ | (16) | after Zilitinkevich et al. (2006a) |
| $C_t = 1$ | (17) | after Zilitinkevich and Baklanov (2002) |

**Figure captions**

Figure 1. Dimensionless velocity gradient, $\Phi_M = \dfrac{kz}{\tau^{1/2}} \dfrac{dU}{dz}$, in ABLs ($z < h$) and above ($z > h$) versus dimensionless height $\xi = z/L_*$, after the LES DATABASE64. Dark grey points show data for $z < h$; light grey point, for $z > h$; the line shows Eq. (11a) with $C_{U1} = 2$.

Figure 2. Same as in Figure 1 but for the dimensionless potential temperature gradient, $\Phi_H = \dfrac{k_T \tau^{1/2} z}{F_\theta} \dfrac{d\Theta}{dz}$. The bold curve shows Eq. (11b) with $C_{\Theta 1} = 1.6$ and $C_{\Theta 2} = 0.2$; the thin lines show its asymptote $\Phi_H = 0.2\xi^2$ ($\sim \xi^2$) and the traditional approximation $\Phi_H = 1+2\xi$ ($\sim \xi$).

Figure 3. Gradient Richardson number, Ri, within and above the ABL versus dimensionless height $\xi = z/L_*$, after the NS ABL data from LES DATABASE64 (dark grey points are for $z < h$ and light grey points, for $z > h$) and observational data from the field campaign SHEBA (green points). Heavy black points with error bars (one standard deviation above and below bin-averaged values) show the bin-averaged values of Ri after the DATABASE64. The bold curve shows Eq. (10) with $\Phi_H$ taken after Eq. (11b), $C_{U1} = 2$, $C_{\Theta 1} = 1.6$ and $C_{\Theta 2} = 0.2$; the steep thin line shows its asymptote: Ri $\sim \xi$; and the thin curve with a plateau (obviously unrealistic) shows Eq. (10) with the traditional, linear approximation of $\Phi_H = 1+2\xi$.

Figure 4. The wind-velocity characteristic function $\Psi_U = k\tau^{-1/2} U - \ln(z/z_{0u})$ versus dimensionless height $\xi = z/L_*$, after the LES DATABASE64. Dark grey points show data for $z < h$; light grey point, for $z > h$. The line shows Eq. (13a) with $C_U = 3.0$.

Figure 5. Same as in Figure 4 but for the potential-temperature characteristic function $\Psi_\Theta = k\tau^{-1/2}(\Theta - \Theta_0)(-F_\theta)^{-1} - \ln(z/z_{0u})$. The line shows Eq. (13b) with $C_\Theta = 2.5$.

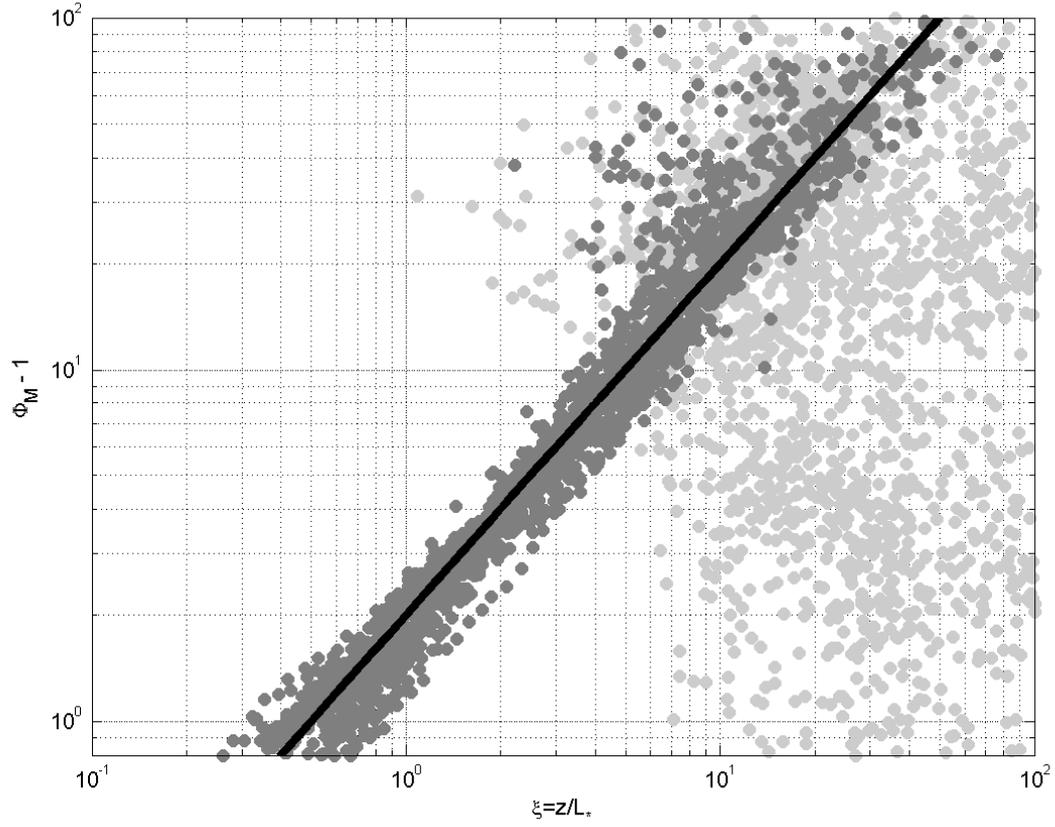

Figure 1. Dimensionless velocity gradient, $\Phi_M = \dfrac{kz}{\tau^{1/2}} \dfrac{dU}{dz}$, in ABLs ($z < h$) and above ($z > h$) versus dimensionless height $\xi = z/L_*$, after the LES DATABASE64. Dark grey points show data for $z < h$; light grey point, for $z > h$; the line shows Eq. (11a) with $C_{U1} = 2$.

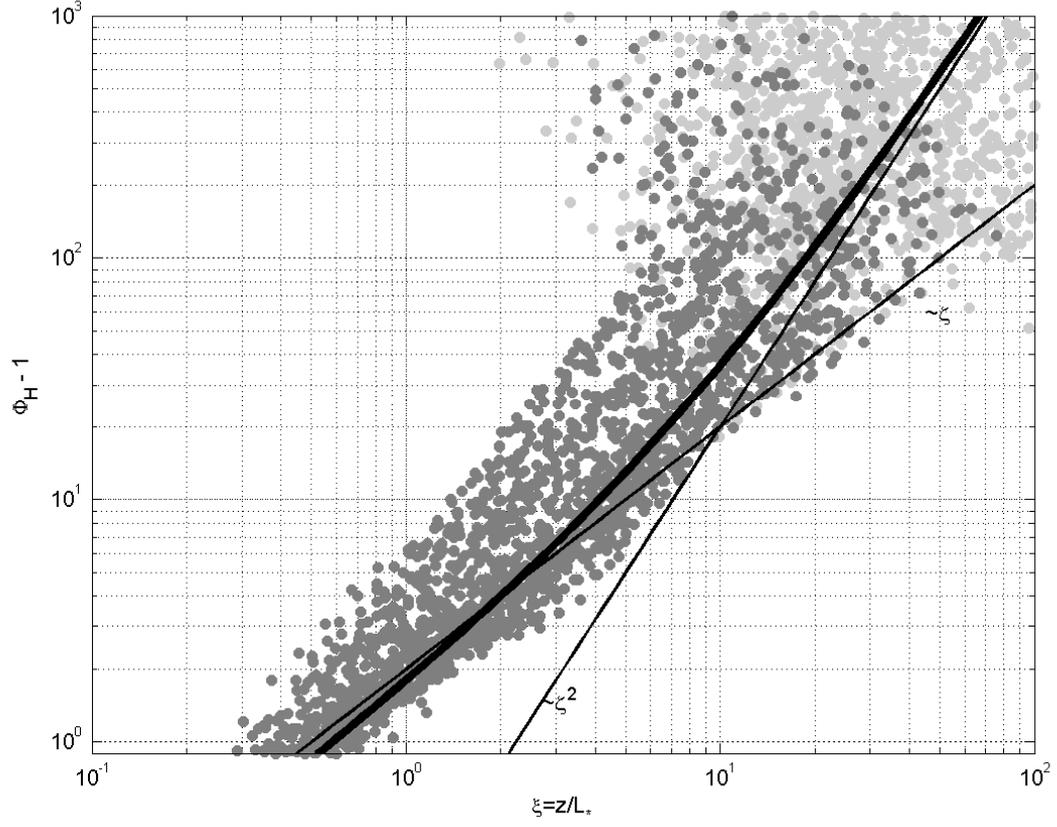

Figure 2. Same as in Figure 1 but for the dimensionless potential temperature gradient, $\Phi_H = \dfrac{k_T \tau^{1/2} z}{F_\theta} \dfrac{d\Theta}{dz}$. The bold curve shows Eq. (11b) with $C_{\Theta 1} = 1.8$ and $C_{\Theta 2} = 0.2$; the thin lines show its asymptote $\Phi_H = 0.2\, \xi^2$ ($\sim \xi^2$) and the traditional approximation $\Phi_H = 1 + 2\xi$ ($\sim \xi$).

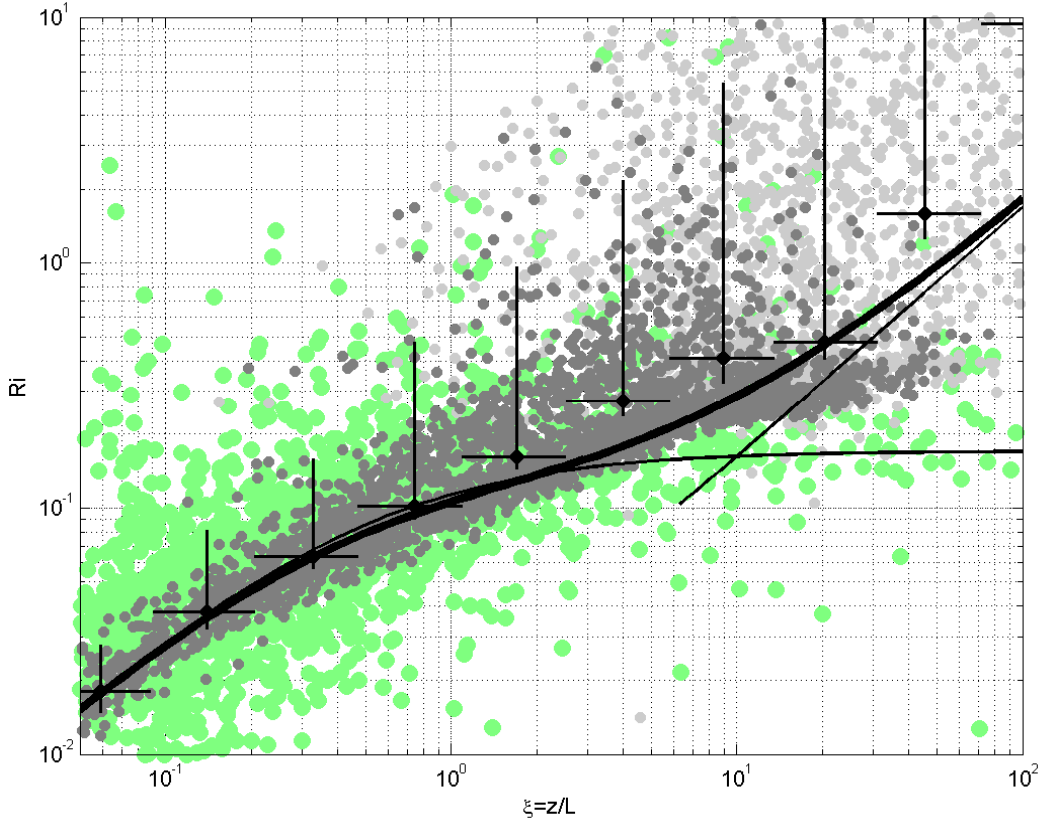

Figure 3. Gradient Richardson number, Ri, within and above the NS ABLs versus dimensionless height $\xi = z/L_*$, after the NS ABL data from LES DATABASE64 (dark grey points for $z<h$ and light grey points for $z>h$) and data from the field campaign SHEBA (green points). Heavy black points with error bars (one standard deviation above and below bin-averaged values) show the bin-averaged values of Ri after the DATABASE64. The bold curve shows Eq. (10) with $\Phi_H$ taken after Eq. (11b), $C_{U1} = 2$, $C_{\Theta 1} = 1.6$ and $C_{\Theta 2} = 0.2$; the steep thin line shows its asymptote: Ri ~ $\xi$; and the thin curve with a plateau (obviously unrealistic) shows Eq. (10) with the traditional, linear approximation of $\Phi_H = 1+2\xi$.

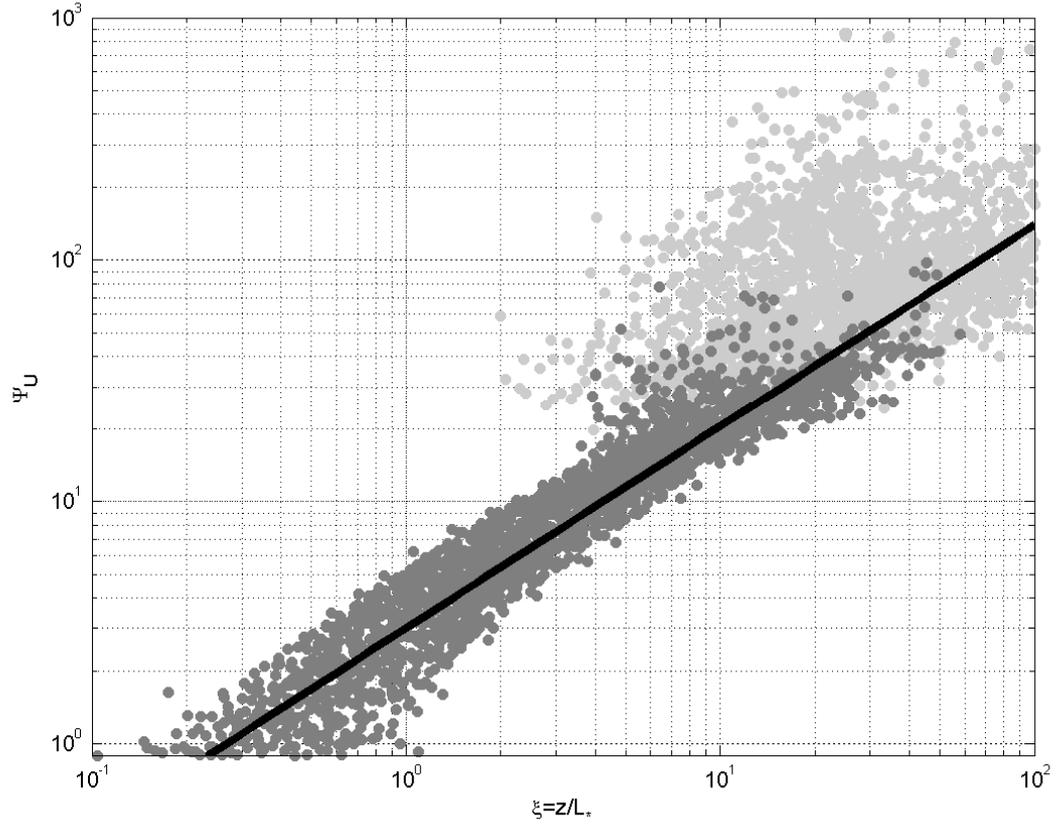

Figure 4. The wind-velocity characteristic function $\Psi_U = k\tau^{-1/2}U - \ln(z/z_{0u})$ versus dimensionless height $\xi = z/L_*$, after the LES DATABASE64. Dark grey points show data for $z < h$; light grey point, for $z > h$. The line shows Eq. (13a) with $C_U = 3.0$.

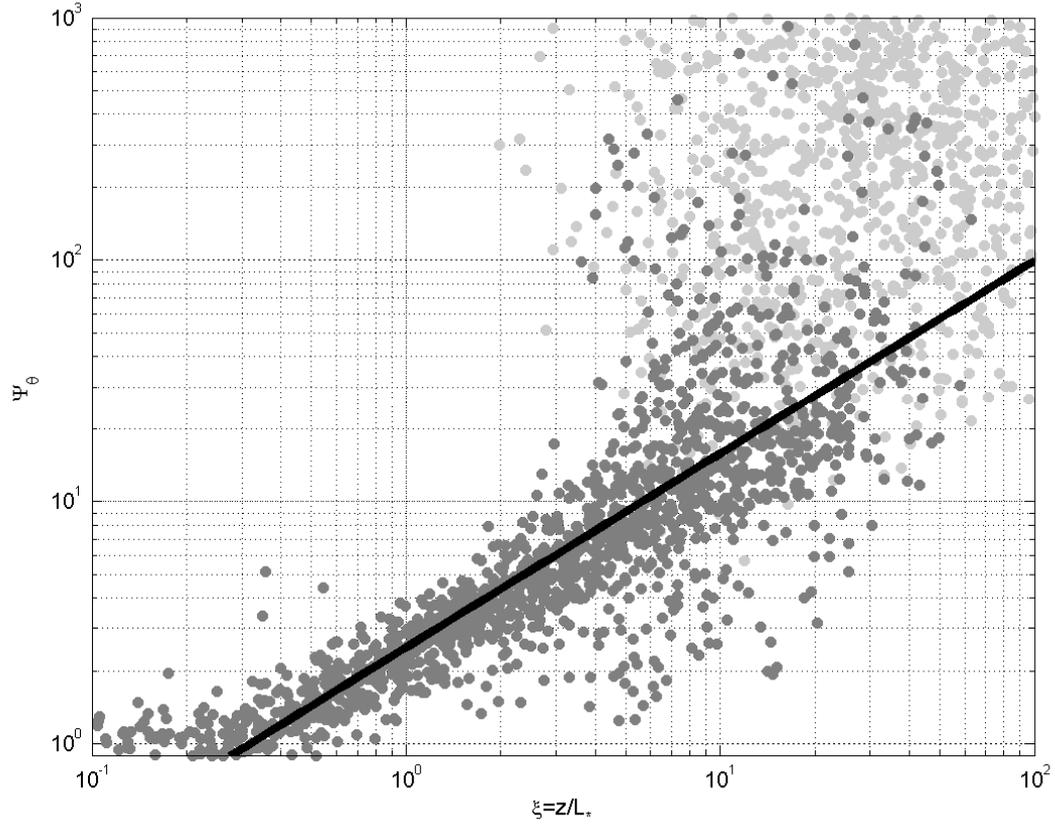

Figure 5. Same as in Figure 4 but for the potential-temperature characteristic function $\Psi_\Theta = k\tau^{-1/2}(\Theta - \Theta_0)(-F_\theta)^{-1} - \ln(z/z_{0u})$. The line shows Eq. (13b) with $C_\Theta = 2.5$.